\documentclass[osajnl2,twocolumn]{revtex4-1}
\usepackage{amsmath}
\usepackage{amssymb}
\usepackage{graphicx}

\begin{document}
\draft \title{Modeling the Mechanisms of the Photomechanical Response of a Nematic Liquid Crystal
Elastomer} \author{Nathan J.
Dawson$^{1,3 \dag}$, Mark G. Kuzyk$^{1}$, Jeremy Neal$^{2}$, Paul Luchette$^{2}$, and Peter Palffy-Muhoray$^{2}$}
\address{$^1$Department of Physics and Astronomy, Washington State University \\ Pullman, Washington  99164-2814 \\
$^2$Liquid Crystal Institute, Kent State University \\ Kent, OH 44242 \\
$^3$Currently with the Department of Physics and Astronomy, Youngstown State University \\ Youngstown, OH 44555 \\
$^\dag$Corresponding author: dawsphys@hotmail.com}
\date{\today}

\begin{abstract}

Recent studies of azo-dye doped liquid crystal elastomers show a strong photomechanical response. We report on models that predict experimental results that suggest photothermal heating is the dominant mechanism in a planar constrained geometry. We compare our models with experiments to determine key material parameters, which are used to predict the dynamical response as a function of intensity. We show that a local strain from photothermal heating and a nonlocal strain from thermal diffusion is responsible for the observed length changes over time. This work both elucidates the fundamental mechanisms and provides input for the design of photomechanical optical devices, which have been shown to have the appropriate properties for making smart materials.

\end{abstract}

\maketitle


\section{Introduction}
\label{sec:introduc}

There has been a growing interest in photomechanical behavior of organic materials over the last decade.\cite{yu03.01,ooste07.01,ooste09.01,dunn09.01,lee10.01} Recently, an experimental study has determined the mechanisms of photo-induced deformations of liquid crystal elastomers in a photomechanical optical device (POD) geometry.\cite{dawson11.03a} This work was motivated in part by the creation of parallel beam PODs made of high-concentration azo-dye doped LCEs that were cascaded together in series to show a proof-of-concept network of sensor/actuator photomechanical devices.\cite{dawson10.01}

The observation of the photomechanical effect dates back to Alexander Graham Bell, who showed that a voice could be transmitted over a beam of light.\cite{ja:bell00.01} A century later, Uchino \textit{et al} demonstrated the ``Uchino walker.''\cite{uchin93.01,uchin90.01,uchin80.01} This device implimented ceramics that constrict when exposed to light, and relax in the dark (reversible photostriction). The mechanism of ceramic photostriction can be explained by light-induced charge diffusion followed by a piezo-electric effect, which is induced by the resulting internal electric field.

With breakthroughs in the technology for making polymer optical fibers, the first all-optical photomechanical device was constructed that acted as a position stabilizer.\cite{welke94.01} This device encompasses five device classes: information processing, logic, transmission, sensing, and actuation.\cite{kuzyk06.06} The next year, a small-scale, $2.5\,$cm, photomechanical optical device was made using an azobenzene-dye-doped multi-mode polymer optical fiber with frayed ends that act as retroreflectors.\cite{welke95.01} This device was later used to demonstrate all-optical modulation of a probe beam by introducing a pump beam that changed the fiber length.\cite{welke96.01} A decade later, a fiber cantilever was observed by the use of off-center propagation of light through a disperse red 1 dye-doped polymer optical fiber.\cite{bian06.01,kuzyk06.06}

Recently, nematic liquid crystal elastomers (LCE) have been reported to undergo large deformations via photo-isomerization.\cite{corbe09.01,Finke01.01} A high-concentration azobenzene-dye-doped nematic LCE was shown to swim over the surface of water when illuminated by a transverse light source.\cite{camac04.01} Other studies of nematic LCEs have investigated the nonlinear behavior of a LCE cantilever.\cite{corb07.01}

Motivated by the recent experimental observations of a nematic LCE in a POD geometry, this work seeks to build an understanding of the mechanisms. We present numerical modeling results of the mechanisms and compare the results with experimental observations of the time-dependent changes in length of a LCE in response to light. Understanding these mechanisms is the first step in the creation of ultra-smart materials.\cite{NloSourceSmart}

\section{Photomechanical Optical Device Configuration}
\label{sec:experiment}

We develop a theory that describes the mechanisms of photo-induced deformations of liquid crystal elastomers in the surface-constrained geometry as described below. During these experiments, a LCE is pressed between two parallel glass substrates with a small force, which is applied to achieve a visible ``wet spot" at the LCE-glass interface. Silver is deposited on the inner surfaces of the glass - away from the position of the LCE. The pump beam is steered into the glass substrate and incident upon one end of the LCE. The probe beam is directed through the interferometer region of the POD, i.e. where the silver is deposited.

\begin{figure}[t!]
\includegraphics[scale=1]{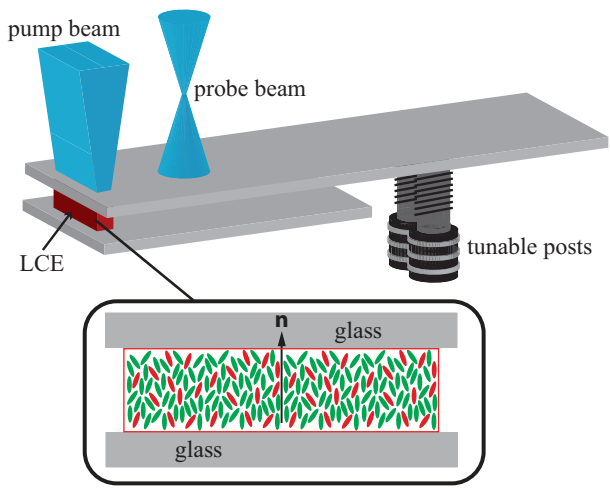}
\caption{(Color online) Diagram of the photomechanical optical device (POD) used to measure the LCE's length change by detecting changes in the interference pattern of the probe beam.}
\label{fig:podconfig}
\end{figure}

\begin{figure}[b!]
\includegraphics[scale=1]{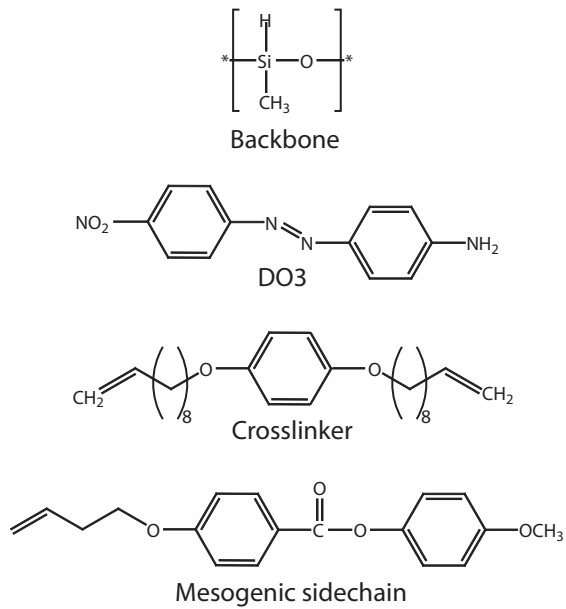}
\caption{The chemical structures of the silicon backbone, trifunctional crosslinker, mesogenic sidechain, and disperse orange 3 dopant chromophore that are used to construct the dye-doped LCEs.}
\label{fig:chem}
\end{figure}

The uniaxial nematic LCE is oriented between the two glass substrates so that the director orientation is normal to the LCE-glass interface. The LCE acts as one of the three stabilizing points that define the planar surface of one substrate. The other two points are defined by the tips of adjustable thumb screws. These were used to bring the two substrates into parallel alignment. Figure \ref{fig:podconfig} shows a perspective diagram of the POD used in previous experiments.\cite{dawson11.03a}

The LCE material within the POD consists of a silicon backbone, crosslinker, and mesogenic sidechain whose structures are shown in Figure \ref{fig:chem}. The sidechain is connected to the backbone by the crosslinker, and there are roughly $10$ silicon backbone segments per crosslinker. Disperse orange 3 (DO3) is a photo-isomerable dye that is dissolved in the LCE to act as an absorber of the pump beam. These absorbers can both change shape and transfer heat to the surrounding mesogens to reduce the macroscopic orientational order.\cite{warne05.01}

Our models treat the high concentration limit, i.e. approximately $0.1\%$ by weight DO3-doped LCEs. The LCEs are pumped with light at a frequency of $488\,$nm, near the resonant absorption peak.\cite{dawson11.03a} At this concentration and wavelength, $\mu\left(Q=0.7\right)\approx 0.2 \,\mu\mbox{m}^{-1}$, where $\mu$ is the Beer-Lambert coefficient (defined in the next section) and $Q$ is the scalar uniaxial order parameter.

\begin{figure}[t!]
\includegraphics[scale=1]{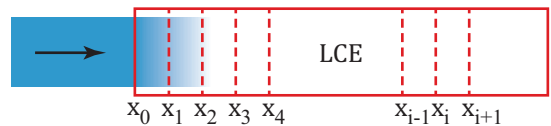}
\caption{(Color online) A liquid crystal elastomer illuminated by a $488$nm wavelength laser, where the direction of light propagation is parallel to the director orientation. The gradient represents laser light absorption.}
\label{fig:figure1}
\end{figure}

\section{Theory}
\label{sec:theorymech}

\begin{figure}[b!]
\includegraphics[scale=1]{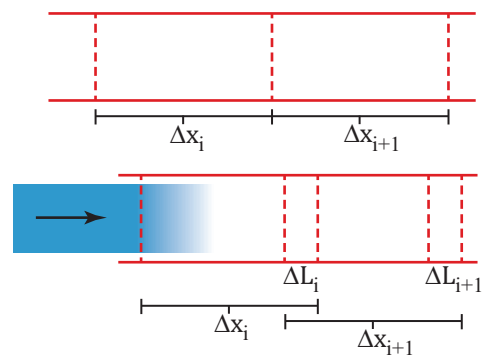}
\caption{(Color online) Sections of an LCE (top) before and (bottom) after turning on the laser. The sections decrease by a length, $\Delta L_i$, at position $x_i$.}
\label{fig:figure2}
\end{figure}

It is well known that a continuous differential equation can be approximated by finite differences provided that the finite pieces are smaller than the smallest characteristic dimension. Figure \ref{fig:figure1} shows a schematic diagram of an illuminated LCE in one dimension.  Each mathematically discrete section of the LCE experiences a length change that is much smaller than the length of each section, which allows for an engineering strain approximation. A diagram is shown in Figure \ref{fig:figure2}.

\subsection{Photo-isomerization induced strain}

The LCE's length at coordinate $x_i$ changes by an amount $\Delta L_i$ depending on the local light intensity and temperature, leading to a strain $\Delta L_i/\Delta x$. For the un-elongated elastomer, the Beer-Lambert law gives
\begin{equation}
    I = I_0 e^{-\epsilon c A l} ,
\label{eq:eqno1}
\end{equation}
where $I$ is the transmitted intensity, $I_0$ the incident intensity, $\epsilon$ the absorption coefficient, $A$ the illuminated area, $c$ the concentration, and $l$ the path length. Figure \ref{fig:figure1} shows that the intensity is maximum at the surface $x=x_0$.

We first consider the effects of photo-isomerization, which is a local mechanism that results in a length change of element $i$ in proportion to the intensity in element $i$. It is well known that the conformation of an azo-dye can change from trans to cis by photo-isomerization.\cite{cvikl02.01} When azobenzene molecules are dissolved into a LCE, the process of photo-isomerization can cause a reduction in the long-range orientational order.\cite{barb00.01}

The probability that an azo-dye in the LCE undergos trans-cis photo-isomerization is proportional to the population of trans isomers oriented along the light's polarization axis and the light's intensity.\cite{Finke01.01} We define $\xi$ as the probability per unit of intensity per unit of time that a trans isomer will absorb light and be converted to a cis isomer. The temperature-dependent decay rate of the cis population, $\beta$, is assumed to be independent of intensity. As long as the temperature increase in the elastomer is small, which it is in our experiment, this will be a good approximation.

The photothermal heating and photo-isomerization mechanisms are coupled because the heating rate depends on the trans population, which depends on the rate of photo-isomerization. If the LCE is highly-doped, so that the 1/e absorption length is much smaller than the length of the sample, only the region near the surface of the sample will be illuminated. Thus, photo-isomerization-induced strain will be limited to a negligibly small part of the sample, where even a large local strain would produce only a small change in the total length of the LCE.

Since the absorption coefficient decreases as trans isomers are converted to cis isomers, the laser penetrates deeper into the sample over time. The equilibrium penetration depth will be determined by the equilibrium trans population, which occurs when the photo-isomerization rate balances the recovery rate, and the difference in optical absorption coefficient of the cis and trans isomer. We find that this penetration depth, the $1/e$ absorption length, is approximately $5 \, \mu \mathrm{m}$ for the range of intensities used in our experiments at a wavelength of $488\,$nm. As such, we can treat the illuminated region as a thin heat source that contributes little to the total length change.

The temperature of the rest of the LCE will increase due to thermal diffusion of heat from the illuminated region, which leads to a reduction in the long-range orientational order parameter away from the optical heat source, and thus a change in the length. By virtue of the fact that the LCE is much longer than the illuminated region, even a small strain in the dark region of a highly azobenzene-doped LCE can lead to a length change that is much larger than in the illuminated region.

Thus, the physical picture of the process is as follows. The light is absorbed in a thin layer of the LCE near its surface. The heat generated in this thin region flows through the sample, causing a decrease in the orientational order of the liquid crystalline molecules due to the resulting change in temperature. The order reduction can be explained using statistical models of nematic liquid crystals such as the method described by Maier and Saupe.\cite{maier58.01,maier59.01,maier60.01}

The change in orientational order causes a length change.  Note that the population of cis isomers of the azobenzene dopants slightly increases with increased temperature, which at \textit{extremely} high concentrations of photonematogen dopants, could also contribute to the decrease of orientational order of the liquid crystal, and thus also contribute to length contraction.\cite{corbe09.01} We lump these two processes together and call it simply the thermal mechanism. The sample reaches thermal equilibrium when the heat flowing out of the LCE into the glass substrates and lost due to convection at the glass-air interface is in balance with the heat generated in the thin surface layer.

Under the assumption that the temperature change is small enough in the illuminated region to invoke the decoupled mechanism approximation, the population dynamics are modeled by light-induced depletion of trans isomers and constant-rate recovery.
The rate of change in the population fraction of trans isomers, $N$, throughout the LCE via photo-isomerization is
\begin{equation}
    \frac{dN\left( x,t \right)}{dt} = - \xi
N\left(x,t\right) I\left(x,t\right) + \beta \left( N_{\mathrm{eq}} - N\left(x,t\right) \right) ,
\label{eq:eqno3}
\end{equation}
where $N_{\mathrm{eq}}$ is the equilibrium population fraction of trans isomers under ambient conditions and $I\left(x,t\right)$ is the light intensity as a function of depth and time. While an increase in the temperature can both cause conversion from cis to trans as well as trans to cis isomers, in equilibrium, the population of the higher-energy species will increase with temperature.

The intensity as a function of depth at the entry surface of the LCE is time-independent, that is, $I\left(0,t\right) = I_0$. Substituting the Beer-Lambert law into Equation \ref{eq:eqno3} gives the nonlinear rate equation
\begin{eqnarray}
& &\frac{d}{d t} N\left(x,t\right) = - \xi N\left(x,t\right) I_0 \exp \left[
-\mu \displaystyle\int_{0}^{x} N\left(x',t\right)dx' \right] \nonumber \\
    & & + \beta \left( N_{\mathrm{eq}} - N\left(x,t\right) \right)
\label{eq:eqno7}
\end{eqnarray}
where $\mu = \epsilon c A$ is the absorption coefficient multiplied by the linear molecular density.

Equation \ref{eq:eqno7} is a generalization of the two state photo-isomerization rate equation in the small absorption limit given by Bian \textit{et al} for photo-reorientation,\cite{bian06.01}
\begin{equation}
    \frac{dN}{dt} = - \xi I N + \beta \left( 1 - 2 N \right),
\label{eq:eqno8}
\end{equation}
which applies to the case when the material is much shorter than the absorption length. The $2N$ term accounts for molecules being equally distributed along two orthogonal directions in a plane perpendicular to the beam polarization axis.

The strain induced by photo-isomerization, $\sigma_p$, is
\begin{equation}
    \sigma_p = \frac{\Delta L_p}{L_0},
\label{eq:eqno9}
\end{equation}
where $\Delta L_p$ is the change in length caused by photo-isomerization and $L_0$ is the initial length.

The order parameter for a LCE is a function of the isomer concentration and temperature. The equation governing the order parameter as a function of trans isomers is calculated self consistently.\cite{barb00.01} With a small cis isomer population at a given depth within the LCE, the uniaxial scalar order parameter, $Q$, can be approximated as a linear function of the dopant concentration, $N$, or $\Delta Q \propto \Delta N$, where $\Delta$ denotes the change in a variable. Also, the curve defining the strain as a function of change in the order parameter will be locally linear over small changes in order. Thus, the strain can then be expressed as $\sigma_t \left(x,t\right) \propto \Delta Q$. This approximation allows the photo-isomerization strain to be expressed as
\begin{equation}
    \sigma_p \left(x,t\right) = - b\left(N_{\mathrm{eq}}-N\right),
\label{eq:eqno10}
\end{equation}
where $b$ is the proportionality constant of photo-isomerization-induced strain.

\subsection{Temperature induced strain}

The other photomechanical mechanism that we consider is photothermal heating.\cite{kuzyk06.06} The temperature distribution throughout the LCE is described by the heat equation. The one-dimensional heat equation with a heat source and conduction-type boundary conditions is given by
\begin{eqnarray}
    \frac {d}{d t} T\left(x,t\right) - K_{\mathrm{LCE}} \frac{d^2}{d x^2} T\left(x,t\right) =
H_{\mathrm{s}}\left(x,t\right) , \label{eq:eqno11}\\
    k_{\mathrm{LCE}} \left.\frac{d}{d n_x} T\left(x,t\right)\right|_{x=0} = f_1\left(t\right) , \label{eq:eqno12}\\
    k_{\mathrm{LCE}} \left.\frac{d}{d n_x} T\left(x,t\right)\right|_{x=L_0} = f_2\left(t\right) , \label{eq:eqno13}
\end{eqnarray}
and
\begin{equation}
T\left(x,0\right) = g\left(x\right) , \label{eq:eqno14}
\end{equation}
where $K_{\mathrm{LCE}}$ is the LCE's thermal diffusivity, $k_{\mathrm{LCE}}$ is the LCE's thermal conductivity, $T$ is the temperature, $H_{\mathrm{s}}\left(x,t\right)$ is the rate of temperature increase due to absorption of light in the material, $f$ describes the temperature gradient as a function of time at each boundary, $n_x$ is the unit vector component along the $\hat{x}$ direction, and $g$ is the initial temperature profile of the material for a LCE of length $L_0$.

Solutions to the boundary value problem depend on the experimental constraints and material parameters. For a steady initial state with no heat source, the temperature profile is constant. Thus, $g = T_0$ (room temperature). Since the change in temperature is of interest rather than the absolute temperature, we subtract the ambient temperature from all temperatures, thereby defining $T_0 = 0$. The light is turned on at $t=0$ to a constant intensity $I_0$, and remains so indefinitely. This step function intensity will result in transient behavior until the system is in equilibrium. A fit of the transient behavior to the model that follows will allow the material's photomechanical parameters to be determined.

The absorption of light from the laser acts as the heat source, so the amount of heating at a point in the LCE is proportional to the light intensity at that point. The boundary function $f$ will be described by contact conduction at the LCE-glass interface and the contact conductance coefficient will be approximated as constant over small temperature increases. Therefore, the boundary conditions may be written as
\begin{eqnarray}
& & \left.\frac{d}{d n_x} T_{\mathrm{LCE}}\left(x,t\right)\right|_{x=0} = \nonumber \\
& & \frac{C_1}{k_{\mathrm{LCE}}} \left[T_{\mathrm{LCE}}\left(0,t\right) - T_{\mathrm{glass}}\left(0,t\right)\right]
\label{eq:bc1first}
\end{eqnarray}
and
\begin{eqnarray}
& & \left.\frac{d}{d n_x} T_{\mathrm{LCE}}\left(x,t\right)\right|_{x=L_0} = \nonumber \\
& & \frac{C_2}{k_{\mathrm{LCE}}} \left[T_{\mathrm{LCE}}\left(L_0,t\right) - T_{\mathrm{glass}}\left(L_0,t\right)\right] .
\label{eq:bc2first}
\end{eqnarray}
Assuming that both boundaries have the same contact conductance coefficient at the LCE-glass interface ($C_1 = C_2 = C$), the heat equation with the above boundary conditions and the initial condition become
\begin{eqnarray}
& &    H_{\mathrm{s}}\left(x,t\right) = \nonumber \\
& & \frac {d}{d t} T_{\mathrm{LCE}} \left(x,t\right) - K_{\mathrm{LCE}} \frac{d^2}{d x^2} T_{\mathrm{LCE}}
\left(x,t\right) , \label{eq:eqno15}\\
& & \left.\frac{d}{d n_x} T_{\mathrm{LCE}} \left(x,t\right)\right|_{x=0} = \nonumber \\
& & \frac{C}{k_{\mathrm{LCE}}}\left[ T_{\mathrm{LCE}}\left(0,t\right)- T_{\mathrm{glass}}\left(0,t\right)\right] , \label{eq:eqno16}\\
& & \left.\frac{d}{d n_x} T_{\mathrm{LCE}} \left(x,t\right)\right|_{x=L_0} = \nonumber \\
& & \frac{C}{k_{\mathrm{LCE}}}\left[ T_{\mathrm{LCE}}\left(L_0,t\right)- T_{\mathrm{glass}}\left(L_0,t\right)\right] ,
\label{eq:eqno17} \\
& & T_{\mathrm{LCE}} \left(x,0\right) = 0 . \label{eq:eqno18}
\end{eqnarray}

The light energy absorbed at any point in the material is proportional to the intensity at that specific point at that time and is also proportional to the number of absorbers oriented along the polarization of the light at that point in the LCE. We define $\zeta$ to be the constant of proportionality that relates the heating rate, $H_{s}$, to the intensity and trans population fraction,
\begin{eqnarray}
    H_{\mathrm{s}}\left(x,t\right) = \zeta N\left(x,t\right) I\left(x,t\right) .
\label{eq:eqno19}
\end{eqnarray}
Given the intensity profile through an absorbing medium, the intensity absorbed is given by,
\begin{equation}
    d I \left(x,t\right)= -\mu I\left(x,t\right) N\left(x,t\right) dx ,
\label{eq:eqno20}
\end{equation}
which can be rewritten as
\begin{equation}
    I\left(x,t\right) N\left(x,t\right) = - \frac{1}{\mu} \frac{d}{d x}
I\left(x,t\right) .
\label{eq:eqno21}
\end{equation}
Substituting Equation \ref{eq:eqno21} into equation \ref{eq:eqno19}, and defining $\alpha = -\frac{\zeta}{\mu}$ yields,
\begin{equation}
    H_{\mathrm{s}}\left(x,t\right) = -\alpha \frac{d}{d x} I \left(x,t\right) .
\label{eq:eqno22}
\end{equation}

Using Equation \ref{eq:eqno22}, Equation \ref{eq:eqno15} can then be written as
\begin{eqnarray}
& & -\alpha \frac{d I\left(x,t\right)}{d x} = \nonumber \\
& & \frac {d}{d t} T_{\mathrm{LCE}} \left(x,t\right) - K_{\mathrm{LCE}} \frac{d^2}{d x^2} T_{\mathrm{LCE}}
\left(x,t\right) , \label{eq:eqno24}
\end{eqnarray}
which can be used to determine fully the thermal spatial strain profile as a function of time of the LCE.

In previous experiments, shown in Figure \ref{fig:podconfig}, the LCE is sandwiched between two glass slides. As such, heat will flow through the glass, then the surrounding air. Given the large volume of glass, heat will flow from the LCE to the glass for the duration of the experiment while the laser is on, and heat convection will occur across the surface of the glass-air interface. This effect will be included in the LCE heat equation to give a more accurate description of the LCE's transient temperature profile.

We assume that the one-dimensional heat equation approximates the sample and the glass slide with contact conduction boundary conditions at the illuminated LCE-glass interface and convection at the other end of the glass slide. Therefore, Equations \ref{eq:eqno15} through \ref{eq:eqno18} will describe the temperature for $0<x<L_0$, and the heat equation for the glass substrates with the step function intensity described earlier will be modeled using the following equations for $-L_g<x<0$.

\begin{eqnarray}
& & \frac {d}{d t} T_{\mathrm{glass}} \left(x,t\right) = K_{\mathrm{glass}} \frac{d^2}{d x^2}
T_{\mathrm{glass}} \left(x,t\right) , \label{eq:eqglass1} \\
& & \left.\frac{d}{d n_x} T_{\mathrm{glass}} \left(x,t\right)\right|_{x=-L_g} =
\frac{h}{k_{\mathrm{glass}}}T_{\mathrm{glass}}\left(-L_g,t\right) , \label{eq:eqglass2} \\
& & \left.\frac{d}{d n_x} T_{\mathrm{glass}} \left(x,t\right)\right|_{x=0} = \nonumber \\
& & \frac{C}{k_{\mathrm{glass}}}\left[
T_{\mathrm{glass}}\left(0,t\right)-T_{\mathrm{LCE}}\left(0,t\right)\right] , \label{eq:eqglass3} \\
& & T_{\mathrm{glass}} \left(x,0\right) = 0 , \label{eq:eqglass4}
\end{eqnarray}
where $k_{\mathrm{glass}}$ is the thermal conductivity, $K_{\mathrm{glass}}$ is the thermal diffusivity, and $L_g$ is the thickness of the glass, while $h$ is the convection coefficient between glass and air at room temperature. Note that the glass has only limited interactions with any substance other than the LCE and air. Equations \ref{eq:eqglass1} through \ref{eq:eqglass4} will also be used to find the temperature throughout the far glass slide at the dark interface located at $L_0<x<L_0+L_g$, where $L_0$ is the coordinate of the LCE-glass interface and $L_0+L_g$ is the glass slide's surface that is in contact with air and cools via convection.

Once the temperature is known throughout the LCE, the strain induced by the change in temperature can be found. In the case of a nematic LCE, the scalar uniaxial order parameter is also temperature dependent. Thus, in the $1/e$ absorption length, photo-isomerization could dominate the photomechanical effect. However, the heat generated in this region can flow to the dark region, leading to a thermal photomechanical response.

The thermal strain, $\sigma_t \left(x,t\right)$, is defined as
\begin{equation}
    \sigma_t = \frac{\Delta L_t}{L_0} ,
\label{eq:straint}
\end{equation}
where $\Delta L_t$ is the change in length caused by the thermal processes.

Similarly, following the previous approximations for photo-isomerization-induced strain, the thermally-induced strain can then be written as $\sigma_t \left(x,t\right) \propto \Delta Q$ where $\Delta Q \propto \Delta T$ for small changes in temperature. Therefore,
\begin{equation}
\sigma_t (x,t) = - q\left[T\left(x,t\right) - T\left(x,0\right)\right]
\label{eq:straintemp}
\end{equation}
where $q$ is a thermal strain proportionality constant (in the same manner as $b$ in Equation \ref{eq:eqno10}).

\subsection{Length change}

The total strain in the decoupled approximation is just the sum of the two strains caused by photo-reorientation and thermal expansion,
\begin{equation}
    \sigma \left(x,t\right) = \sigma_t \left(x,t\right) + \sigma_p \left(x,t\right) .
\label{eq:eqno26}
\end{equation}

The strain in Equation \ref{eq:eqno26} is then used to evaluate the total change in length of the LCE. The change in length is related to the average strain, $\sigma_{\mathrm{avg}}$, by
\begin{equation}
    \sigma_{\mathrm{avg}} \approx \frac{\Delta L}{L_0} \hspace{0.7cm} \mbox{for}
\hspace{0.7cm} \Delta L \ll L_0 ,
\label{eq:eqno27}
\end{equation}
where the total length change is $\Delta L = \Delta L_p + \Delta L_t$. If the strain were uniform, then the total change in length would be the strain multiplied by the initial length. In contrast, in our experiments the strain depends on the depth, so that the change in length in each segment of the LCE depends on its location. Therefore, if the initial length of each segment is $\Delta x$, then the change in length of the $i^{th}$ segment is $\Delta L_i$, which is given by,
\begin{equation}
    \Delta L_i = \sigma(x_i) \Delta x .
\label{eq:eqno28a}
\end{equation}
The total length change can be calculated by adding up the length of each section,
\begin{equation}
    \Delta L = \displaystyle\sum_{i=0} \sigma(x_i) \Delta x .
\label{eq:eqno28b}
\end{equation}
In the continuum limit, if the strain is sufficiently small, the total length can be approximated by
\begin{equation}
    \Delta L = \displaystyle\int_{0}^{L_0} \sigma(x,t) dx .
\label{eq:eqno29}
\end{equation}

Over a long time period, the LCE reaches an equilibrium strain profile, accompanied by an equilibrium population profile, temperature profile, and intensity profile. A measurement of $\Delta L$ as a function of time during the transition to equilibrium can be used to determine the phenomenological parameters from Equation \ref{eq:eqno29} with the solutions obtained for Equation \ref{eq:eqno26} which is used as the input of the numerical approximations to Equation \ref{eq:eqno24}.

\subsection{Reversibility of LCE length change}

The laser is turned off after the system reaches equilibrium. Once again, the heat equation is solved but with no heat source; and, the initial strain profile for the relaxation process will be simply the strain the instant the laser is turned off.

The photo-isomerization-induced strain is treated in the same manner, where the population fraction of trans isomers along the direction of the polarization of the electric field becomes
\begin{equation}
    \frac{d}{d t} N(x,t) = \beta \left( N_{\mathrm{eq}} - N(x,t) \right) ,
\label{eq:eqno34}\end{equation}
where the initial condition is just the population fraction of absorbers throughout the LCE at the time the laser is turned off.

The heat equation for the glass substrates are identical to the previous ones except that the initial condition is now the temperature profile at the time the laser is turned off. Similarly, the temperature profile of the LCE at laser shut-off must be used as the initial conditions when solving for the LCE's relaxation. The equations governing the cooling of the LCE are
\begin{eqnarray}
& & \frac {d}{d t} T_{\mathrm{LCE}} \left(x,t\right) = K_{\mathrm{LCE}} \frac{d^2}{d x^2} T_{\mathrm{LCE}}
\left(x,t\right) , \label{eq:eqno35} \\
& & \left.\frac{d}{d n_x} T_{\mathrm{LCE}} \left(x,t\right) \right|_{x=0} = \nonumber \\
& & \frac{C}{k_{\mathrm{LCE}}}\left[
T_{\mathrm{LCE}}\left(0,t\right)-T_{\mathrm{glass}}\left(0,t\right)\right] , \label{eq:eqno36} \\
& & \left.\frac{d}{d n_x} T_{\mathrm{LCE}} \left(x,t\right)\right|_{x=L_0} = \nonumber \\
& & \frac{C}{k_{\mathrm{LCE}}}\left[
T_{\mathrm{LCE}}\left(L_0,t\right)-T_{\mathrm{glass}}\left(L_0,t\right)\right] , \label{eq:eqno37} \\
& & T_{\mathrm{LCE}} \left(x,0\right) = G\left(x\right) , \label{eq:eqno38}
\end{eqnarray}
where $G\left(x\right)$ is the initial temperature profile of the LCE the instant the laser is turned off.

As before, these equations can be used to calculate the individual strains due to the decoupled processes and summed. Note that during the relaxation process, we use the same parameters as those determined from the transient response after the laser was turned on.

\section{Computational Results}
\label{sec:numapproxi}

For the case of photothermal heating, we must take into account the heat generated by the laser at the input end as well as thermal diffusion through the sample and glass substrates. The photo-isomerization mechanism in the illuminated region is ignored due to the extremely short path length. Furthermore, because of the small strain observed in previous POD experiments, we will use the engineering strain approximation and will neglect the effects due to the constraint imposed at the LCE-glass interfaces.

We solve the heat equation for the glass and LCE simultaneously using the constraints due to the boundary and the material's constitutive relations to get a set of equations describing thermal-induced length changes using the finite difference approximation. The increments are small enough to obtain a good approximation to our experimental parameters.

The explicit finite difference method was chosen due to the multiple boundaries that are taken into account. The photo-isomerization equation and heat equation were solved in relatively large spatial increments, $66\,\mu$m, and small temporal increments, $1\,$ms, to account for stability governed by the Fourier number, which must be taken into consideration for the stability of the explicit finite difference approximation. Because the path length of the pump beam is aprroximately $5\,\mu$m for a high concentration DO3-doped LCE, all of the light is absorbed in the first LCE increment. Thus, the change in the population fraction of trans isomers is negligible in the heat source term throughout the rest of the sample. This picture is consistent with experimental observations of a photomechanical response that is dominated by thermal diffusion. Using these approximations, the heat equation was numerically solved to determine the temperature profile from which the length change was determined. The numerical models for length contraction and relaxation were used as fitting functions, where $q$ is the thermal strain proportionality constant, and is introduced as a phenomenological fitting parameter.

The thermal strain, $\sigma_t(x,t)$, peaks near the surface of the LCE at short times, and propagates inward as the heat diffuses beyond the illuminated region. Photo-isomerization orientational hole burning leads to a decrease in optical absorption, and thus allows the light to penetrate further into the sample. The light continues to burn an isomerization hole until an equilibrium intensity profile is reached, defined by the condition that the rate of isomerization induced by the laser matches the relaxation rate at each point in the sample. Since experimental evidence suggests that the photo-isomerization's contribution to the strain from the illuminated region is negligible, the step size in the numerical calculations was chosen to be much larger than the $1/e$ absorption length to avoid the necessity for modeling the effects of orientational hole burning.

Since photo-isomerization is found to occur in a thin surface layer of a high concentration DO3-doped LCE (typically less than $50\,\mu$m), the contribution to the length change of the LCE from this mechanism is negligible. The dynamics of the photo-isomerization process, however, determines the temporal dependence of energy absorption; but, once this fast process reaches equilibrium, its net effect can be approximated as a constant heat source near the sample's surface. Thus, the mechanisms of photo-isomerization and heating are effectively decoupled.

Previous work supports the view that length contraction due to photo-isomerization is negligible.\cite{dawson11.03a} This approximation will remain valid as long as the light is absorbed over a thin-enough length that its contribution to the length change is negligible. In this regime, we can set $N \approx N_\mathrm{eq}$, and
\begin{equation}
\sigma\left(x,t\right) \approx \sigma_t \left(x,t\right).
\label{eq:totstrain}
\end{equation}

The length change of the LCE can be determined from the angle between the plates and the length change at the position of the probe laser using a simple geometric analysis based on the geometry of similar triangles. An estimate of the initial phase of the interferometer at the start of the experiment is the largest source of experimental uncertainty. The positions of the peaks and the initial phase of the interferometer provides sufficient information to determine the evolution of the length of the LCE during and after laser illumination.

To determine the length change, we use the transmittance equation for a Fabry-Perot interferometer,
\begin{equation}
    \frac{I\left( t \right)}{I_{\mathrm{max}}} = \frac{1}{1+F \sin^2 \left(
\frac{2\pi}{\lambda}\left(\Delta L\left(t\right) + L_0\right) \right)} ,
\label{eq:eqno35}
\end{equation}
where $F$ is a constant, commonly known as the finesse, that depends on the reflectivity of the mirrors. For a $488\,$nm pump laser, the separation between peaks corresponds to a displacement of $244\,$nm.

The theory is used to fit the data for a range of laser powers of high concentration dye-doped LCEs. A few simplifying assumptions were made. First, the thermal diffusivity is assumed to be constant, independent of temperature and degree of photo-isomerization of the dyes in the LCE. Also assumed is that the contact conductance coefficient through the LCE-glass interface remains constant and is equal on both the illuminated interface and the opposite dark interface. The convection coefficient, $h$, is also assumed to be a constant. Finally, the beer-lambert coefficient is also assumed to be temperature independent.

\begin{figure}[b!]
\includegraphics[scale=1]{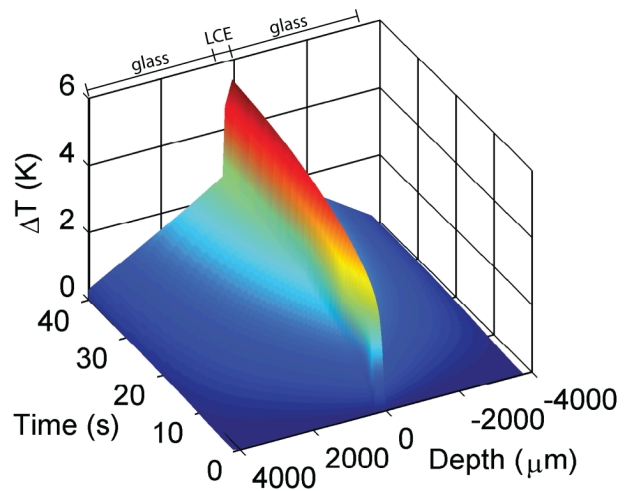}
\caption{(Color online) The calculated temperature change due to heating by absorption of a $36\,$mW laser over the surface of a $400\, \mu$m LCE initially at room temperature. The middle peak corresponds to the position of theLCE while the broader wings correspond to the glass substrate.}
\label{fig:numtemp}
\end{figure}

The results of a numerical simulation of the temperature over a $40$ second interval is shown in Figure \ref{fig:numtemp}, which shows the temperature as a function of time and depth in the LCE and glass slides. Heat flows out through all the surfaces of the $1\,$mm-thick glass substrates through convection. The substrate's large volume makes numerical modeling difficult due to memory and computing power limitations. Our approach is to apply a one-dimensional approximation; but, increasing the thickness out to a distance that is larger than the actual substrate thickness but smaller than its two largest dimensions to take into account the larger volume of the substrate. While the thermal diffusivity of the glass is set to the literature value, the heat transfer coefficient is adjusted so that the average heat transfer through the sample/glass interface approximates the actual value in the presence of the large glass substrates. This is a good compromise because it takes into account heat conductance at the sample surface where the temperature is greatest and approximates the effects of the parts of the substrate that are furthest from the interface, which will not effect the sample's temperature profile significantly. In effect, this approach averages over the effects of the remote parts of the bulk substrate.

\begin{figure}[b!]
\includegraphics[scale=1]{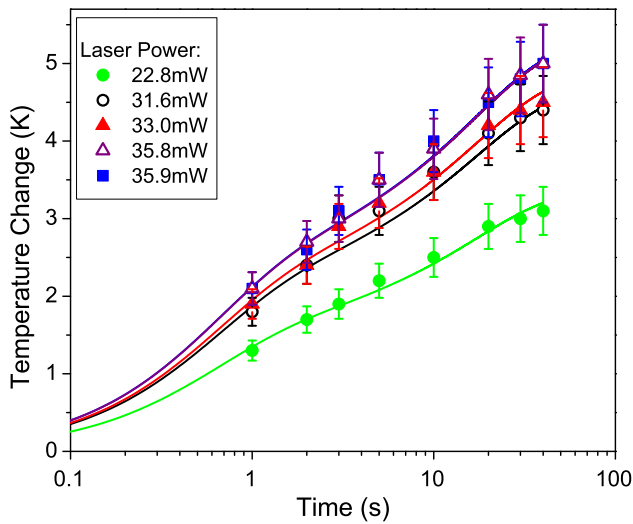}
\caption{(Color online) The change in temperature for an experimental run lasting $40$ seconds (points) and the numerical calculation (curves) using the parameters shown in Table \ref{table:parameters}. The time scale is logarithmic.}
\label{fig:multitemp}
\end{figure}

Figure \ref{fig:multitemp} shows the average temperature from numerical calculations and temperature probe measurements plotted against time. The theory and experiment agree well.

\begin{figure}[t!]
\includegraphics[scale=1]{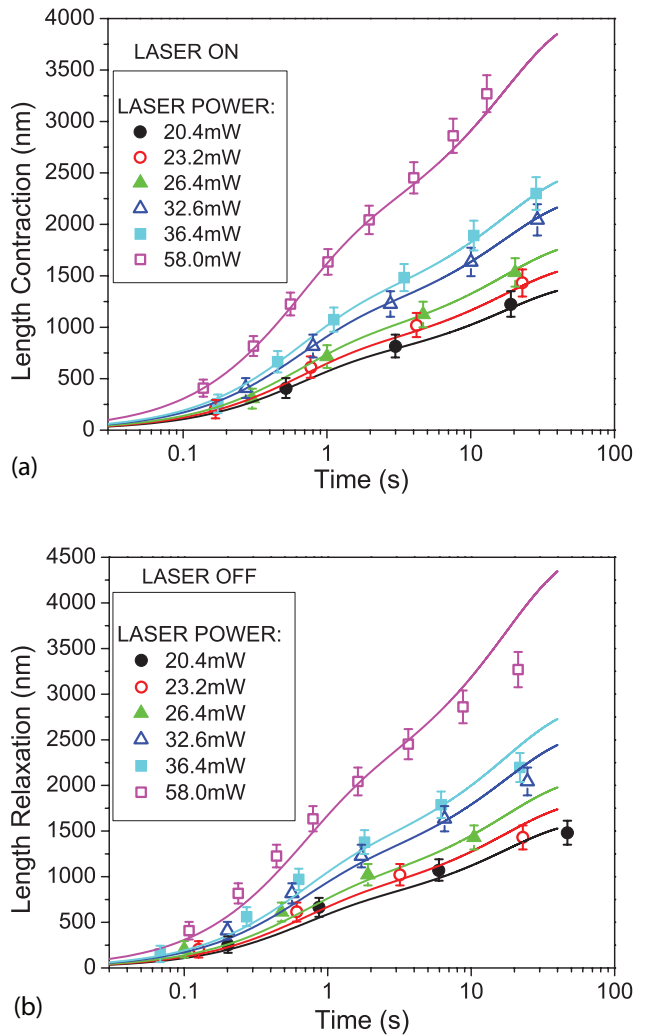}
\caption{(Color online) (a) The length contraction of an LCE for a series of incident laser powers as a function of time after the laser is turned on, and (b) the length relaxation when the laser is turned off after being illuminated for $5$ minutes. The theory curves use the parameters shown in Table \ref{table:parameters}.}
\label{fig:lengthplots}
\end{figure}

The total strain approximated as the thermal strain in Equation \ref{eq:totstrain} is used to model the total length change of the sample, and the numerical results are fit to the experimental data.  The length change data from measurements and the theoretical curves for several intensities are shown in the top of Figure \ref{fig:lengthplots}. Note that the high optical density of the sample leads to high energy density of absorption, so that one must be careful to use intensities that are low enough to prevent damage to the LCEs.

The values of the parameters from the literature or determined from fitting are shown in Table \ref{table:parameters}. Note that the constant of proportionality, $q$, for a dye-doped LCE is about an order of magnitude larger than it is for DR1-doped PMMA (approximately $2.5\times 10^{-4}\,$K$^{-1}$) as reported by Xiang and co-workers.\cite{xiang06.01} This difference is primarily due to the leveraged effect of a photothermally-induced decrease in orientational order of the mesogens.

The parameters from the fits of the LCE length as a function of time in the presence of light were either determined separately by experimentation, found in literature, or determined by fits using the numerical calculation. The data and the fits for the illuminated response are shown in the top graph of Figure \ref{fig:lengthplots}. The bottom graph in Figure \ref{fig:lengthplots} shows the theoretical curves and data of length relaxation to the initial state as a function of time after the light beam is turned off. The theoretical curves showing relaxation use the same parameters as were determined from the light-induced studies. Also note that the set of parameters are the same for all curves at the various light intensities.

The calculation of length relaxation over time after the laser is turned off spans $40$ seconds of time.  Since our model does not take into account the changes in the long-time boundary condition, the theoretical predictions for the total length contraction are always slightly smaller than the experimental values. Note that assumptions about the geometry of the glass and LCE leads to a small deviation between the experimental data and the theory at longer times. However, the predicted trends are consistent with the data.

To elaborate, with one set of parameters, our theoretical model of photo-isomerization decoupled from photothermal heating accurately predicts both the length increase as a function of time of an LCE for a wide range of intensities, as well as the relaxation process (Note that $q$ is a function of the initial nematic order and can change from sample to sample). Thus, the parameters are intrinsic to the system, and their dependence on the laser power is observed to be negligible. This self consistency of a broad set of data that is described by one set of parameters that agree with literature values suggest that our model of the mechanisms of the photomechanical response is valid.

\begin{table}[t!]
\caption{LCE Parameters} 
\centering
\begin{tabular}{l l l l} 
\hline\hline 
Constant & Value & Units & Source\\ [0.5ex] 
\hline
$q$ & $1.181 \times 10^{-3}$ & $\mbox{K}^{-1}$ & determined\\
$K_{\mathrm{LCE}}$ & $1.5 \times 10^{-7}$ & $\mbox{m}^2 \cdot \mbox{s}^{-1}$ & literature/fit\\
$K_{\mathrm{glass}}$ & $4.1 \times 10^{-7}$ & $\mbox{m}^2 \cdot \mbox{s}^{-1}$ & literature\\
$k_{\mathrm{LCE}}$ & $0.195$ & $\mbox{W} \cdot \mbox{m} \cdot \mbox{K}^{-1}$ & literature/fit\\
$k_{\mathrm{glass}}$ & $1.1$ & $\mbox{W} \cdot \mbox{m} \cdot \mbox{K}^{-1}$ & literature\\
$\alpha / A$ & $2.5 \times 10^{-3}$ & $\mbox{W}^{-1} \cdot \mbox{m} \cdot \mbox{K}$ & fit\\
$C$ & $3.575 \times 10^{2}$ & $\mbox{W} \cdot \mbox{m}^{-2} \cdot \mbox{K}^{-1}$ & fit\\
$h$ & $2.75\times 10^{3}$ & $\mbox{W} \cdot \mbox{m}^{-2} \cdot \mbox{K}^{-1}$ & fit\\
$\mu$ & $2.0 \times 10^5$ & $\mbox{m}^{-1}$ & determined\\ [1ex] 
\hline
\end{tabular}
\label{table:parameters}
\end{table}

\section{Conclusion}
\label{sec:concluding}

We have developed a general theory of the photomechanical response of a nematic LCE in a POD geometry. Photothermal heating was able to accurately predict the complicated rate of length change by accounting for the surrounding's thermal properties that a simple rate equation for localized photo-isomerization can not explain. The two mechanisms are coupled through the isomerization hole-burning effect, where the population fraction of trans isomers is depleted, which makes the material less absorbing and allows light to travel deeper into the material. This coupling effect was found to be negligible within the geometries of a POD. The experimental results for high concentrations of photo-isomerizable dye dissolved in a LCE are consistent with the decoupled approximation because most of the energy is absorbed near the surface, which then diffuses throughout the LCE leading to a large thermal-induced strain.

Our model correctly predicts the dynamical behavior of a liquid crystal elastomer in response to a laser that is abruptly turned on, remains at a constant intensity until the system reaches equilibrium, and turned off abruptly. One set of parameters describes both the onset and relaxation of the LCE length over a broad range of intensities up to the point where the material is near its damage threshold. As such, experimental results, when fit to the theory, can be used to measure the material's photomechanical constants in this geometry. Furthermore, the models accurately predict both the temperature change and length change as a function of time, and shows a strong correlation between them. Moreover, this correlation is independent of pump intensity. Therefore, the set of theoretical models that is in concert with experimental results provide strong evidence that photothermal heating is the dominant mechanism.

{\bf Acknowledgements:} MGK and PPM thank the National Science Foundation (ECCS-0756936)
for generously supporting this work. NJD and MGK thank the Air Force Office of Scientific Research (FA9550-10-1-0286) for their generous support.

\end{document}